\documentclass{article}
\usepackage[utf8]{inputenc}
\usepackage{float}
\usepackage{graphicx,csquotes}
\MakeOuterQuote{"}
\usepackage{subfig,hyperref,cite}
\title{Blockchain for Academic Credentials}
\author{Chaitanya Bapat}
\date{23 April 2019}

\begin{document}

\maketitle

\section{Abstract}
Academic credentials are documents that attest to successful completion of any test, exam or act as a validation of an individual's skill. Currently, the domain of academic credential management suffers from large time consumption, high cost, dependence on third-party and a lack of transparency. A blockchain based solution tries to resolve these pain-points by allowing any recruiter or company to verify the user credentials without dependence on any centralized third party. Our decentralized application is based off of BlockCerts, an MIT project that acts as an open standard for blockchain credentials. The project talks about the implementation details of the decentralized application built for BlockCerts Wallet. It is an attempt to leverage the power of the blockchain technology as a global notary for the verification of  digital records.

Keywords - Academic Credentials, Verification, Blockchain, Bitcoin.

\section{Introduction}
Academic credentials include but not restricted to diplomas, degrees, certificates, and certifications, act as a way to attest completion of training or education undertaken by the student. Broadly speaking, these credentials may also attest to successful completion of any test or exam. Ultimately, they serve as a mode of independent validation of the said individual's possession of the knowledge, skill and ability needed to carry out a particular task or activity.
The paper talks about applying Blockchain to the field of education technology (specifically for "Academic credentials")

However, the current scenario of academic credentials suffers from multiple problems. As far as verification of academic credentials is concerned, existing process involves manual work on the part of educational institutions which is time-consuming as well as costly. Besides, this process uses outdated paper credentials which seem unfit in today's digital world. Besides, such a mechanism is not tolerant to a system down-time. Existing process also suffers from the issue of migration. For example students from different countries possess certificates in different languages. To grasp an understanding and comprehensibility of those certificates is at times not possible. Besides, due to natural and man-made calamities, at times schools are shut or cease to exist. During such instances, the certificates and credentials are no longer available. On the flip side, there is a perennial problem of academic fraud. It is mainly classified into three main categories - false degree, fake university, falsified resume. In a nut shell, some chronic, unattended problems exist. Blockchain can prove to be a handy solution for addressing the aforementioned issues.

\section{Motivation}

Assessments may it be for application for higher education, employment, licensing with a professional association or even immigration to another country, is a painstaking task, that is often under-estimated. Besides, different organizations perform assessments in different ways. Moreover, due to lack of standards and requirements, it is subject to interpretation and hence is associated with ambiguity and uncertainty.

Another problem with the existing scenario is lack of decentralization and the current dependence on a trusted third-party. It is partly true that there exist services currently that help enable verification of the credentials called as "credential evaluators". Some of the famous credential evaluators, especially when it comes to United States of America, include but not restricted to  International Education Research Foundation, Inc. (IERF), Academic Evaluation Services (AES), World Education Service (WES), Educational Credential Evaluators, and Inc. (ECE). \cite{international_student} Many a times, there are lot of discrepancies in the evaluations with different evaluators providing different evaluations for the same document. Similarly, there are other platforms like Verifiedcredentials.com \cite{verifiedcredentials} that do similar job of performing background check of indiividuals for businesses and institutions. They also suffer from similar issues of cost, reliance on third party and lack of transparency. All in all, it is not streamlined and involves trusting these third-parties. 

Motivation for applying blockchain to the domain of academic credentials involves 3 actors: students, educational organizations and third party.

Motivation for students:
\begin{itemize}
    \item Document portability\\
    By offloading the credentials from a physical world to an online (digital) world, document portability is achieved to certain extent. But, due to blockchain technology, portability is truly achieved as documents can only be shared to the person who needs to see them; moreover the recipient can verify the existence of the document due to blockchain. In addition, documents are now available anytime and anywhere due to its digitization.
    \item Permanence of records\\
    Many a times, students who possess physical copies of their documents face the risk of loss of records. With blockchain, since it provides immutability, the academic records once hashed onto the blockchain will continue to exist as long as the blockchain exists.
    \item Security and Privacy\\
    Actual data is never stored on the blockchain and only the compressed cryptographic hash of the certificate is put on the blockchain. Moreover, it is identified only on the basis of the address given by the recipient. 
    \item Reduced time and cost of acquisition\\
    Students end up spending a considerable amount of money for acquiring their digital credentials on top of the physical ones and their replacements. Now that physical records would become obsolete, replacement is no longer relevant. Apart from this, the time involved in procuring this document varies from few days to few months. With digitization and blockchain, it will be reduced to a few minutes if not less.
\end{itemize}

For university/education institutions
\begin{itemize}
    \item Brand\\
    Credential from reputed public and private educational organizations/universities, online platforms will all be recognized by their public address. However, as the blockchain adopts "claims based orientation to identification", it gives a chance for issuers to build their corresponding brand. Moreover, since no one else can falsely claim to have attended or received a credential from that organization, institutions would now be able to protect their brand.
    \item Safe Storage\\
    Storage of physical hard copies or digital records of the academic records poses lots of risks for the organizations. By leveraging blockchain's immutability, storage safety can be resolved seamlessly.
    \item Reduced Overhead cost\\
    Overhead costs of an educational organization for managing of academic credentials involves administrative costs such as employee salaries, employer queries, man-hours, document procurement cost and legal costs if any. Leveraging the decentralized architecture of blockchain would offload all these overhead costs dramatically.
    \item Minimized Risk of Loss\\
    With permanence of blockchain credentials, blockchain greatly reduces any risk associated with loss of academic records due to external or human errors introduced as a part of manual handling process.
    \item Environment Sustainability\\
    As every institution, organization and country battles environment pollution and global warming, reduction of carbon footprint in any form is an achievement of sorts. Due to automation of physical copies, tonnes of paper would be saved as re-use of certificates/diplomas is no longer required.
\end{itemize}
For third party organizations
\begin{itemize}
    \item Trustable\\
    Academic fraud is the most complained issue when it comes to verification of academic records of applicants. Most companies spend countless money, man-hours in performing background checks of their employees which involves verifying if their academic records are legitimate. With blockchain based academic credentials, trust on third party verifiers is no longer needed and it becomes much more 
    \item Efficient verification process\\
    Employers will now have instant access to the documents of potential employees allowing them to perform background checks efficiently. This will greatly reduce the time and costs associated with the entire process helping employers as well as potential employees.
\end{itemize}

\section{Literature Survey}
When it comes to academic credential management, there are multiple solutions out there in the market currently. Traditionally speaking, an individual was supposed to take care of his/her own academic credentials which was a painstaking process of applying for transcripts, getting them posted and ensuring they are stored safely and accessible anytime.

\textit{TrustEd}, a platform that is trying to achieve exactly the same thing as our project. It is a platform that leverages blockchain technology for storage, issuance and authentication of academic credentials. It uses Ethereum as the backbone of their platform. Moreover, Ethereum allows them to use smartcontracts and create token of their own. In order to manage the payments and receipts on the \textit{TrustEd} platform, they have developed \textbf{TrustEd Token} (TED) which is ERC20 standard compliant. Usage of \textit{TrustEd Credits} (TCRD) allows the platform to achieve lower transaction fees and higher scalability of operations within the TrustEd network. All in all, the white paper talks in length about the implementation details of the entire platform.

\section{Role of Blockchain}
After understanding the problems in existing system, how does blockchain come into all this? How does it fix all these issues? What power it gives you? Where does security come from that prevents forging or changing your information? All these answers are answered in the following sections as we explain the role of blockchain.

Certificates possess the blockchain property of \textbf{immutability} and hence can't be updated. The official recommendation is to use a logical group name for batches of certificates. Although, they are immutable i.e. the information they contain can't be changed, but in case of any mistake, issuer can simply revoke the entire batch of certificate and re-issue the updated certificates. The immutability of blockchain is lent by its data structure, with each block building upon the last. Compressing the data through hash and logging it into blockchain allows it to be tamper-proof, for all intents and purposes.

It is a common question of how blockchain solves reliance on third party. When there is already a public key infrastructure (PKI) in place, why would one use blockchain? Short answer to that would be \textbf{decentralized trust}. In brief, issuer of the credential uses their own digital signature to send the academic record to the recipient, who is identified by their own public key. Once on the blockchain, the credential of the recipient contains the Merkle proof that can be tied to a specific transaction on the blockchain. In case of PKI, to establish the authenticity of the issuer, time stamping authority (TSA), a trusted third party is required to ensure that the issuer possessed his own private key at the time of issue of credential. In contrast, in case of blockchain, it itself provides a decentralized, permanent, trusted timestamp mechanism by design. 

The blockchain commitment works in a way to ensure verification of credentials is achieved in a secure, distributed and reliable manner. Once issuer issues the credential based on the recipient's public key (blockchain address), it is hashed on the merkle tree that acts as a proof. If this hash were to be forged, it could only be done so by either finding a collision of the hash function, or undoing multiple blocks. Both of these require massive computation power, so the validity of the certificate relies on these, which we assume to be secure, so the certificate is valid. 

Leveraging blockchain technology, also provides added benefits in terms of \textbf{cost and time saving}. BlockCerts being a Bitcoin based blockchain, has it's principle mode of interaction via transactions. A Bitcoin transaction is calculated based on the size of the transaction and the transaction fees. The size of a BlockCerts transaction is static and small, as apart from the standard single input, single output transaction only a small fixed-size \verb|OP\_RETURN|. Thus, regardless of the number of certificates to be issued in a single batch, the cost of the transaction is largely going to be determined by the transaction fee. While a higher transaction fee ensures faster issuance on to the blockchain, the actual dollar value cost differs based on the current blockchain price. However, it is known to be significantly cheaper than issuing hard copies or digital copies of each credential as is the existing process.

As far as privacy of information is concerned, such sensitive and private information is not stored on blockchain and hence not available. Basically, the certificate with the contents is passed through a one-way hash function and that hash is stored on the blockchain. Storage of one-way hash makes it perfectly suitable for application such as third-party verification making it infeasible to recover the original data. Such, a \textbf{cryptographic hash} helps to ensure that privacy of the user is maintained. 

\section{Implementation}
Blockcerts is an open standard for building applications that issue and verify blockchain-based official records. Such applications are called as dApps or "Decentralized applications" as they leverage the distributed ledger technology (blockchain). Blockcerts started in 2016 with the Bitcoin blockchain and then soon expanded to Ethereum. It is an open-source project allowing contributors to continue its development for making Blockcerts work across public as well as private chains. Going forward, it is also being made compatible with the Open Badges, a group of specifications and open technical standards originally developed by Mozilla Foundation with funding from the MacArthur Foundation. The Open Badges standard outlines a method for packaging information about the accomplishments, embedding it into portable images akin to "digital badge" along with an infrastructure for badge validation. According to the Blockcerts roadmap, \cite{blockcerts_roadmap} both the standards will be made compatible with each other and the work is still in progress.\\

\subsection{Working}
This entire process of management of credentials is assumed to start with the \texttt{Issuer} (for example school) sending an invite to all the \texttt{recipients} to receive a blockchain credential. As it stands, we have set it up in the form of an email message. The template for the email message can be found in the code repository and anyone (\texttt{issuer}) who wishes to curate the content of the mail can do so. The next step involves the \texttt{recipient} accepting the invitation sent over email and then as a response, the \texttt{recipient} should send his/her blockchain address to the issuer. Once the \texttt{issuer} has the required blockchain address of the \texttt{recipient}, it hashes the credential onto the blockchain. Bear in my mind that in order to save up on the storage space, the entire certificate is never put on the blockchain. Only the hash of the certificate along with the corresponding blockchain address is put on the blockchain. Moreover, to save up on the storage space, the transactions can be bundled together in one block. Logically speaking, if the \textit{class of 2019} graduating students need to be conferred the \textit{Masters degree}, \textbf{Georgia Tech} could create transactions for all the graduating Masters students together in one block. Once the credentials are hashed, \texttt{issuer} sends the \texttt{recipients} their blockchain credential. This blockchain credential can be easily viewed on the client-side decentralized application. Moreover, such a blockchain credential can be shared with any other third-party by the \texttt{recipient}. It is important to note that the ability to share the blockchain credential rests with the \texttt{recipient}. \texttt{Third party} can use the blockchain to verify the credential.
\begin{figure}[H]
\centering
\includegraphics[scale=0.5]{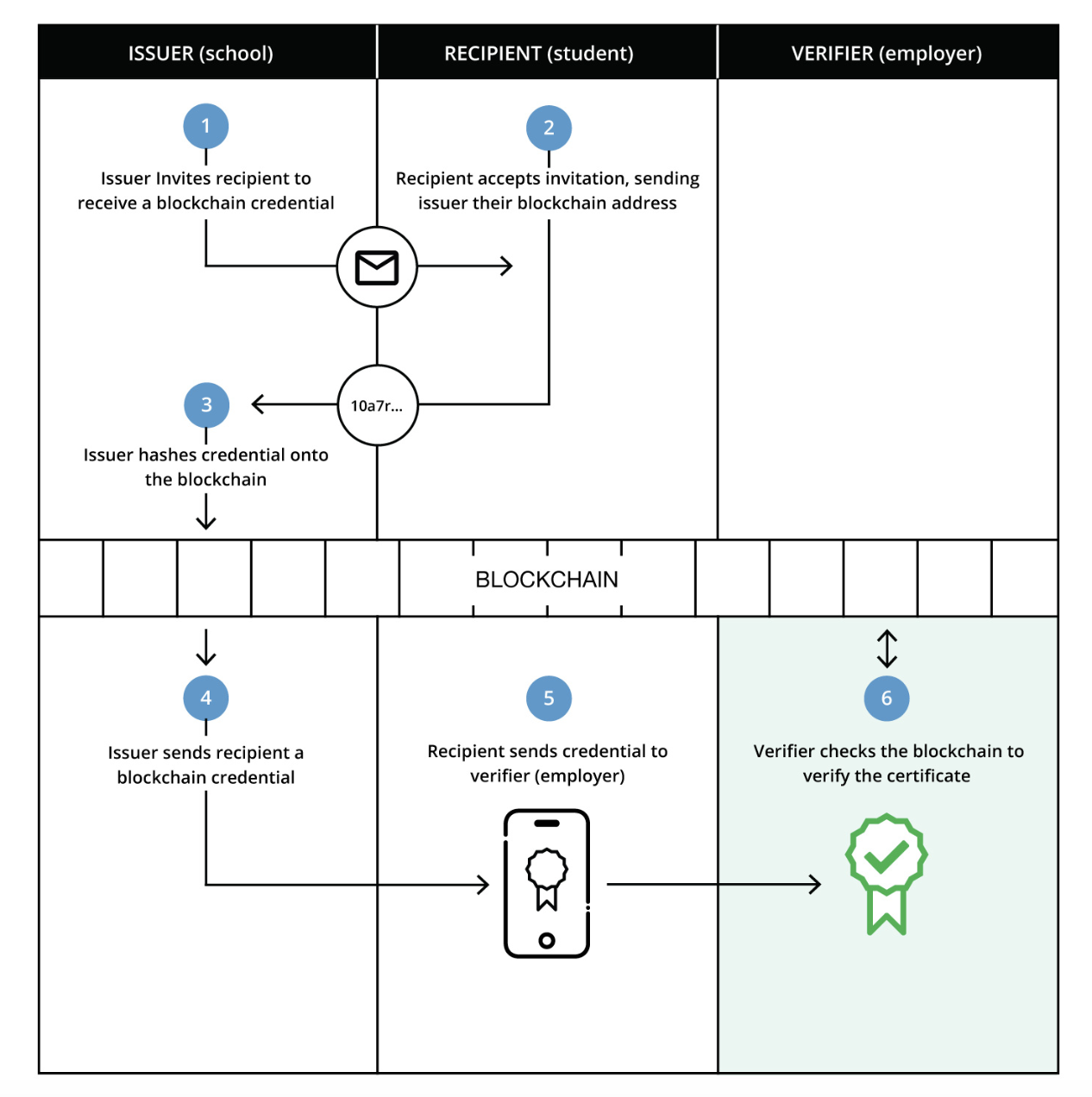}
\caption{How Blockcerts works?}\label{flowchart_blockcert}\cite{blockcerts}
\end{figure}
\subsection{Verification}
\begin{itemize}
    \item \textbf{Format Validation}: Performs all the primary sanity checks to ensure the formats meet the requirement.
    \begin{itemize}
        \item Get TX id: Fetches the transaction identifier.
        \item Compute local hash: Uses the transaction data to pass it through the one-way hash function
        \item Fetch remote hash: Retrieves the hash stored onto the blockchain
        \item Get issuer profile: Fetches the profile of the issuer of the blockchain credential
        \item Parse issuer keys: Reads the issuer keys used for issuing the credential
    \end{itemize}
    \item \textbf{Hash Comparison}: Involves actual verification.
    \begin{itemize}
        \item Compare hashes: Verifies if the two hashes (local and remote) are equal.
        \item Check Merkle root: Ensures that merkle root has not changed i.e. data hasn't been tampered.
    \end{itemize}
    \item \textbf{Status Check}
    \begin{itemize}
        \item Check Revoked status: Verifies if the credential issued to the recipient has been revoked by the issuer.
        \item Check Expiry Date: Verifies if the credential issued by the issuer has passed its expiry date if it exists.
    \end{itemize}
\end{itemize}

\begin{figure}[H]
\centering
\includegraphics[scale=0.1]{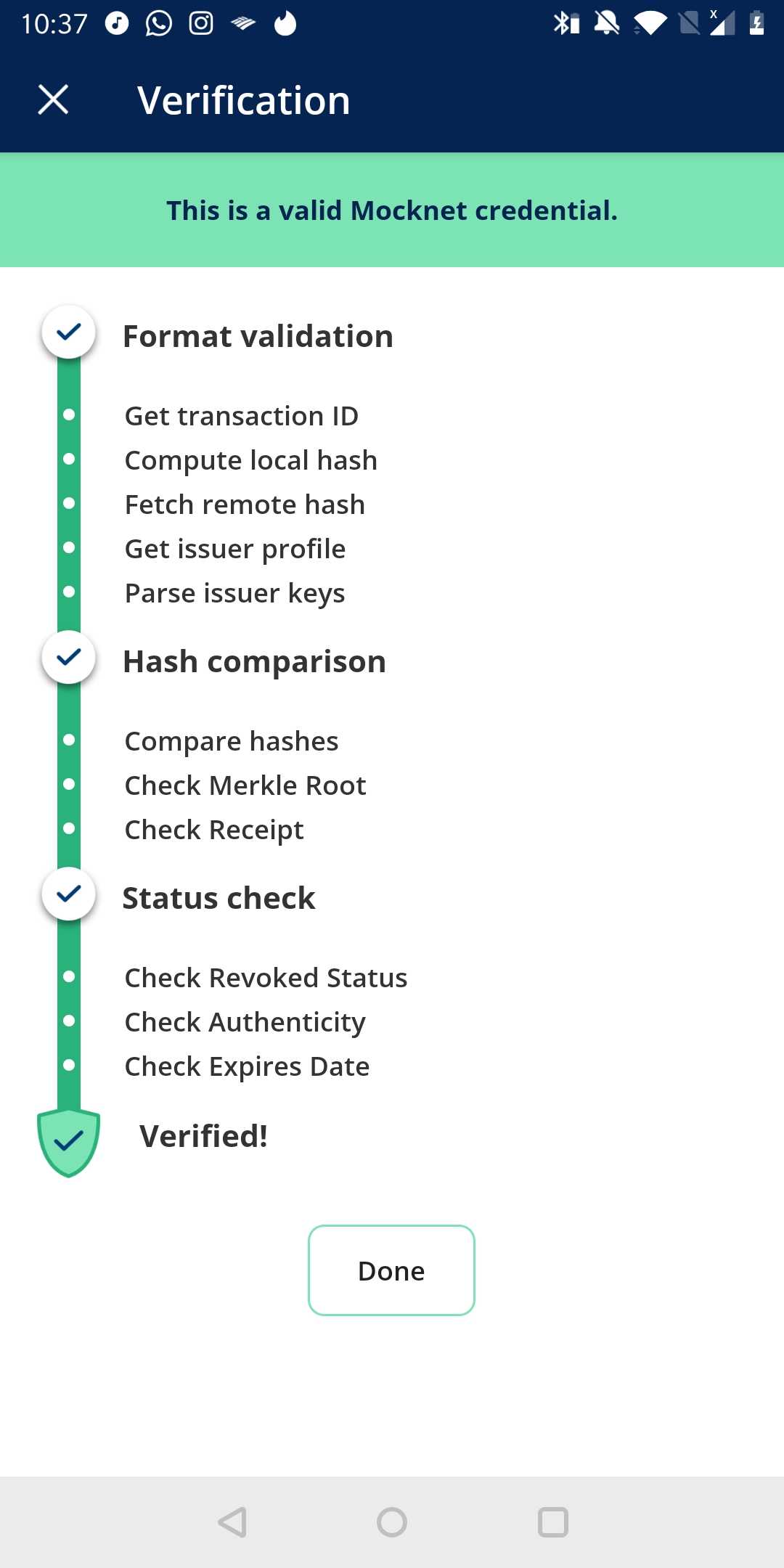}
\caption{Verification of BlockCerts}\label{verify_blockcert}
\end{figure}
Figure ~\ref{verify_blockcert} : Verification tab outlines all the steps carried out to verify the particular credential as it will appear to any \texttt{third-party}.

\subsection{Mock Ups}
\begin{figure}[H]
  \centering
  \subfloat[Home Screen]{\includegraphics[width=0.24\textwidth]{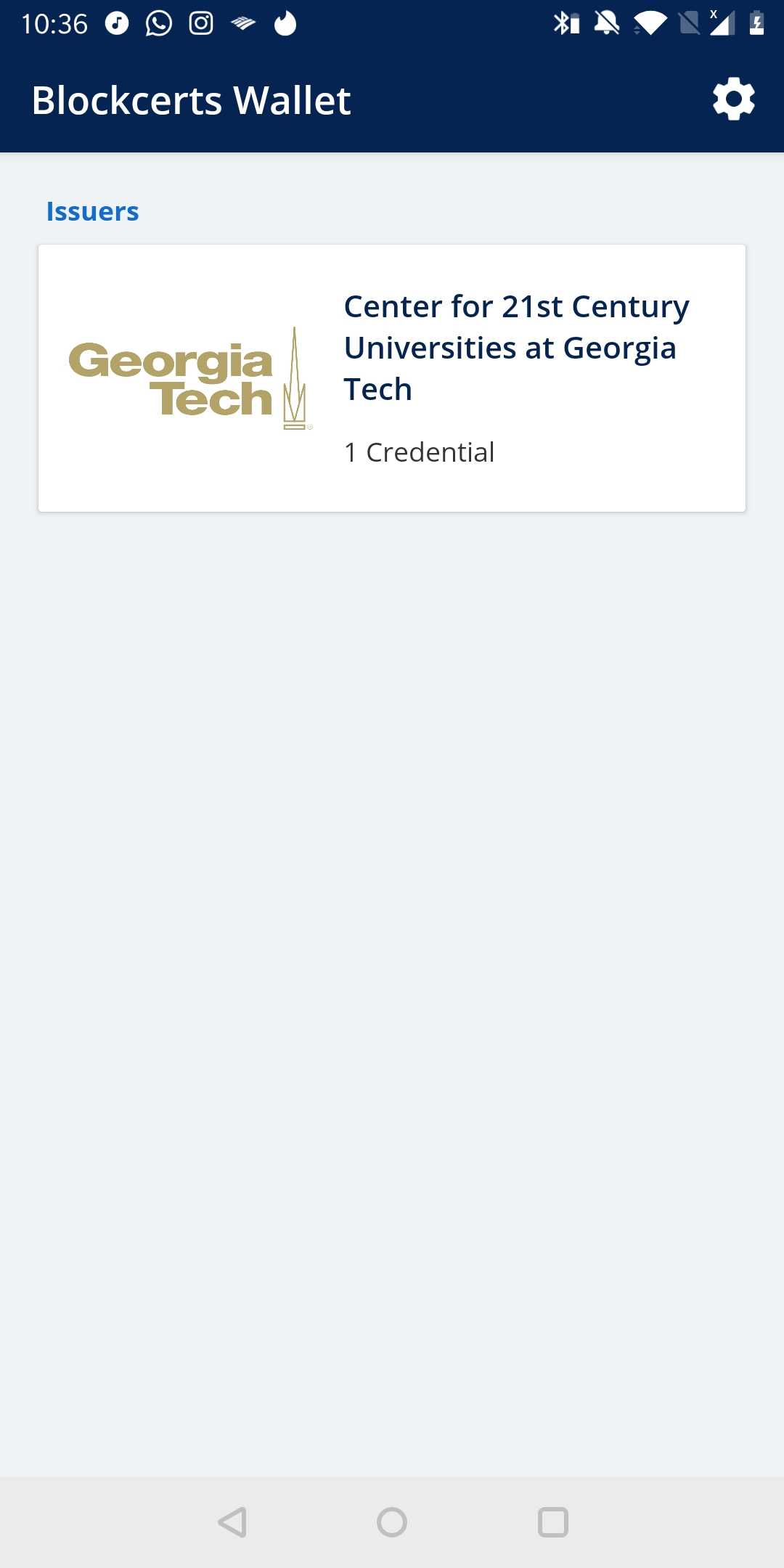}\label{fig:f1}}
  \hfill
  \subfloat[Issuer specific]{\includegraphics[width=0.24\textwidth]{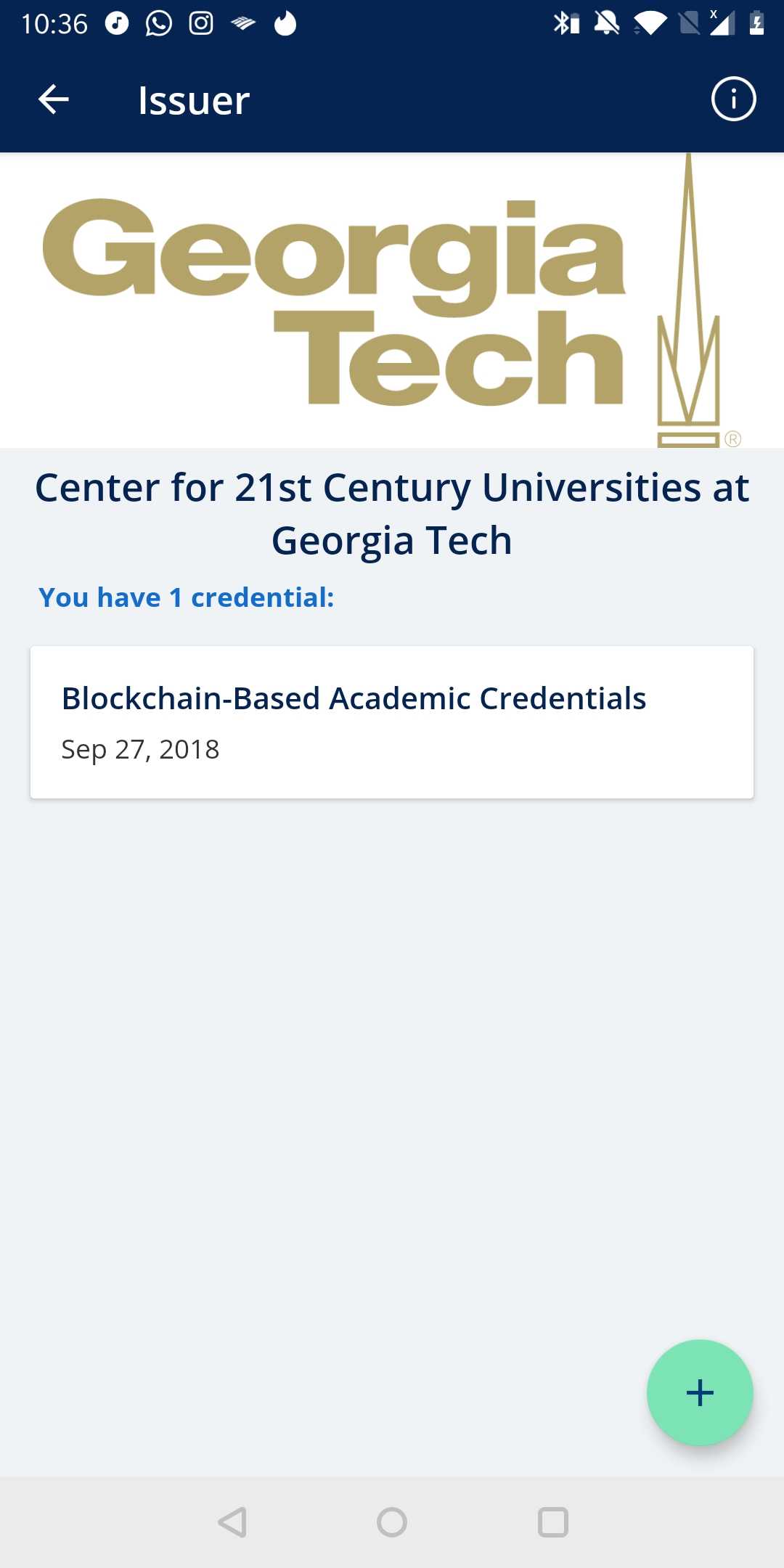}\label{fig:f2}}
  \hfill
  \subfloat[Dummy Credential]{\includegraphics[width=0.24\textwidth]{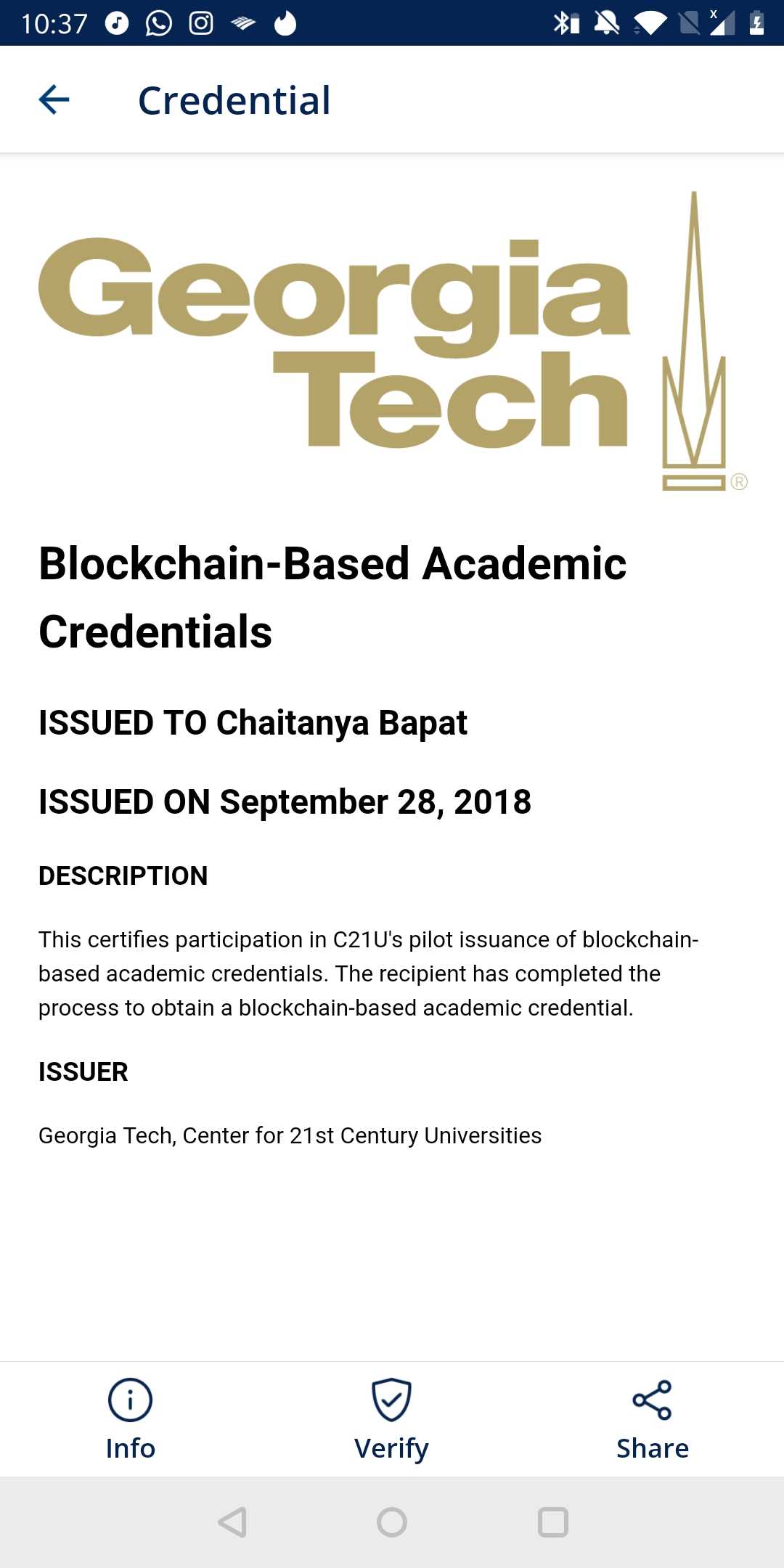}\label{fig:f3}}
  \hfill
  \subfloat[Credential Info]{\includegraphics[width=0.24\textwidth]{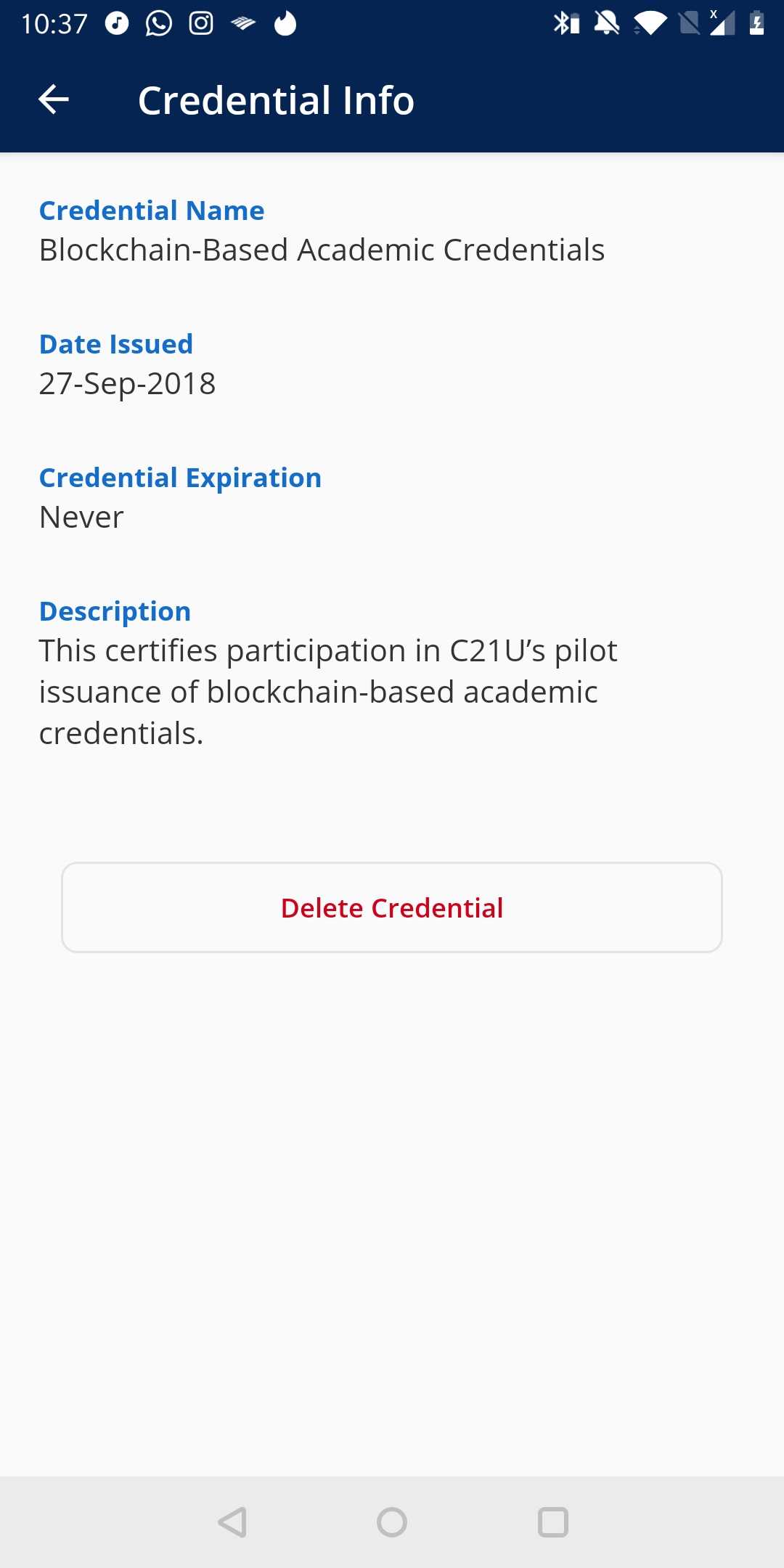}\label{fig:f4}}
  \caption{Mock-up Screens of the Recipient Client dApp}
\end{figure}

\begin{itemize}
    \item Figure ~\ref{fig:f1} : Home Screen displays the BlockCerts Wallet. It shows all the \texttt{issuers} from whom credentials have been earned by the \texttt{recipient}.  
    \item Figure ~\ref{fig:f2} : \texttt{Issuer}-specific screen contains all the credentials received from this particular \texttt{issuer} by the user 
    \item Figure ~\ref{fig:f3} : Dummy credential shows how the actual credential will appear on the application with all the relevant details mentioned.
    \item Figure ~\ref{fig:f4} : Credential Info contains meta-data about the credential.
\end{itemize}
\section{Conclusion}
The paper highlights some chronic problems present in the existing scenario for management of academic credentials. To address the issues, the paper tries to come up with a cryptographically secure, decentralized and trust-less solution via blockchain. In addition, the paper outlines the advantages and motivation for leveraging blockchain technology in this case. Moreover, an attempt has been made to demonstrate the application of BlockCerts for issue, receipt and sharing of a blockchain-based academic credential. Going further, it can be applied to all sorts of credentials, not just academic.
\bibliography{ref.bib}{}
\bibliographystyle{plain}
\end{document}